\documentclass[preprint]{revtex4-1}  
\usepackage{graphicx}  
\usepackage{natbib}
\usepackage{tabularx}

\usepackage{amsmath}
\usepackage{amssymb}

\newcommand{\be}{\begin{equation}}

\newcommand{\ee}{\end{equation}}
\newcommand{\bes}{\begin{eqnarray}}
\newcommand{\ees}{\end{eqnarray}}

\newcommand{\figref}[1]{Fig.~\ref{#1}}
\renewcommand{\eqref}[1]{Eq.~(\ref{#1})}

\begin{document}

\title{Automated, predictive, and interpretable inference of {\em C.~elegans} escape dynamics}

\author{Bryan C.\ Daniels}
\affiliation{ASU--SFI Center for Biosocial Complex Systems, Arizona State
  University, Tempe, AZ 85281, USA}

\author{William S.\ Ryu}
\affiliation{Department of Physics and The Donnelly Centre, University of
  Toronto, Toronto, ON M5S 1A7, Canada}

\author{Ilya Nemenman}
\affiliation{Department of Physics, Department of Biology, and Initiative
  in Theory and Modeling of Living Systems, Emory University, Atlanta,
  GA 30322, USA}






\begin{abstract}
The roundworm {\em C.~elegans} exhibits robust escape behavior in response to rapidly rising temperature. The behavior lasts for a few seconds, shows history dependence, involves both sensory and motor systems, and is too complicated to model mechanistically using currently available knowledge. Instead we model the process phenomenologically, and we use the {\em Sir Isaac} dynamical inference platform to infer the model in a fully automated fashion directly from experimental data. The inferred model requires incorporation of an unobserved dynamical variable, and is biologically interpretable. The model makes accurate predictions about the dynamics of the worm behavior, and it can be used to characterize the functional logic of the dynamical system underlying the escape response. This work illustrates the power of modern artificial intelligence to aid in discovery of accurate and interpretable  models of complex natural systems. \end{abstract}


\maketitle

The quantitative biology revolution of the recent decades
has resulted in an unprecedented ability to measure dynamics of
complex biological systems in response to perturbations with the
accuracy previously reserved for inanimate, physical systems. For
example, the entire escape behavior of a roundworm {\em Caenorhabditis
  elegans} in response to a noxious temperature stimulus can be
measured for many seconds in hundreds of
worms~\cite{Mohammadi:2013ku,LeuMohRyu16}. At the same time,
theoretical understanding of such living dynamical systems has lagged
behind, largely because, in the absence of symmetries, averaging, and
small parameters to guide our intuition, building mathematical models
of such complex biological processes has remained a very delicate art.
Recent years have seen emergence of {\em automated modeling}
approaches, which use modern machine learning methods to automatically
infer the dynamical laws underlying a studied experimental system and
predict its future dynamics
\cite{Schmidt:2009dta,Schmidt:2011gf,SusAbb09,Neuert:2013ed,DanNem15,DanNem15b,Brunton:2016ds,Mangan:is,Lu:2017fy,Pathak:2018dg,Pandarinath:2017bu,HenHemFra18}.
However, arguably, these methods have not yet been applied to {\em any} real
experimental data with dynamics of {\em a priori} unknown structure to
produce interpretable dynamical representations of the system. Thus
their ability to build not just statistical but {\em
  physical} models of data \cite{nelson}, which are interpretable by a
human, answer interesting scientific questions, and guide future
discovery, remains unclear.

Here we apply the {\em Sir Isaac} platform for automated inference of
dynamical equations underlying time series data to infer a model of the
{\em C.\ elegans} escape response, averaged over a population of
worms. We show that {\em Sir Isaac} is able to fit not only the
observed data, but also to make predictions about the worm dynamics
that extend beyond the data used for training. The inferred optimal
model is fully interpretable, with the identified interactions and the
inferred latent dynamical variable being biologically meaningful. And by analysing the dynamical structure of the model---number of dynamic variables, number of attractors (distinct behaviors), etc.---we can generalize these results across many biophysical systems.

\section*{Results} 

\subsection*{Automated Dynamical Inference} 

{\em Sir Isaac} \cite{DanNem15b,DanNem15} is one of the new generation
of machine learning algorithms able to infer a dynamical model of time
series data, with the model expressed in terms of a system of
differential equations. Compared to other approaches, {\em Sir Isaac}
is able to infer dynamics (at least for synthetic test systems) that
are (i) relatively low-dimensional, (ii) have unobserved (hidden or
latent) variables, (iii) have arbitrary nonlinearities, (iv) rely only
on noisy measurements of the system's state variables, and not of the
rate of change of these variables, and (v) are expressed in terms of
an interpretable system of coupled differential equations. Briefly, the
algorithm sets up a complete and nested hierarchy of nonlinear
dynamical models. {\em
  Nestedness} means that each next model in the hierarchy is more
complex (in the sense of having a larger explanatory power)
\cite{vapnik-book,rissanen-book,Bialek:2001wv} than the previous one,
and includes it as a special case. {\em Completeness} means that {\em
  any} sufficiently general dynamics can be approximated arbitrarily
well by some model within the hierarchy. Two such hierarchies have
been developed, one based on S-systems \cite{Savageau:1987vt} and the
other on sigmoidal networks \cite{sigmoidal}. Both progressively add
hidden dynamical variables to the model, and then couple them to the
previously introduced variables using nonlinear interactions of
specific forms. {\em Sir Isaac} then uses a semi-analytical
formulation of Bayesian model selection \cite{MacKay:1992ul,Balasubramanian:1997vr,Bialek:2001wv,DanNem15b} to choose the model in the hierarchy that best balances
the quality of fit versus overfitting and is, therefore,
expected to produce the best generalization.
The sigmoidal network hierarchy is especially well-suited to
modeling biological systems, where rates of change of variables
usually show saturation, and it will be the sole focus of our study.

\begin{figure*}[t]
\centering
\includegraphics[width=6.5in]{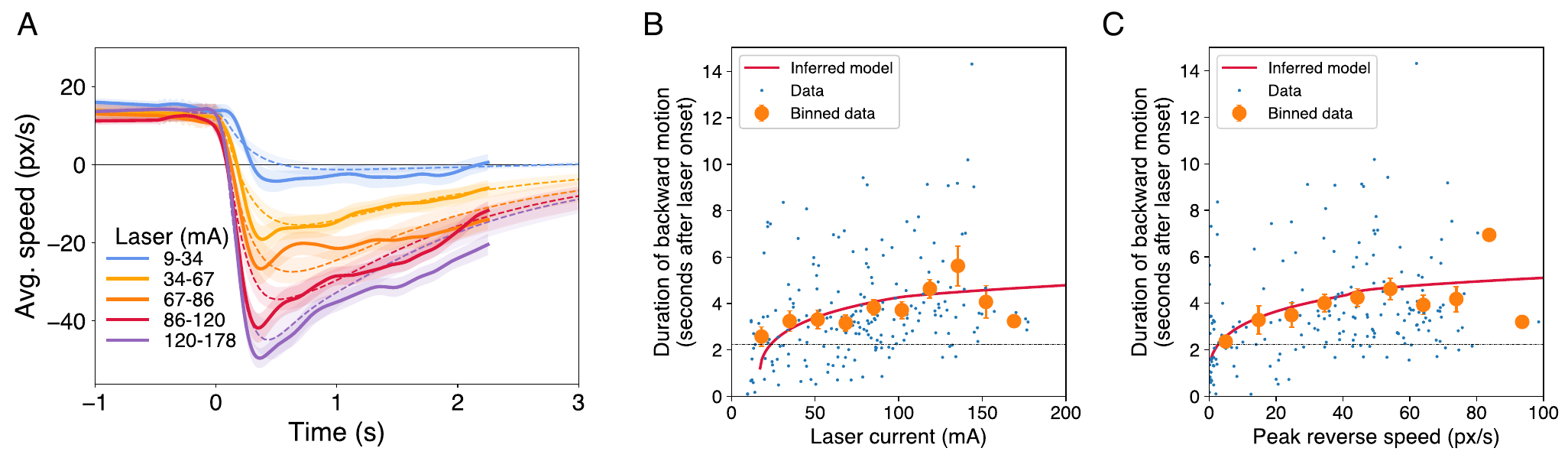}
\caption{ {\bf The escape response behavior is fitted and predicted
    well by the inferred model.} (A) Colored lines and shaded bands
  represent the empirical mean and the standard deviation of the mean,
  respectively, of the escape response velocity for five groups of
  worms stimulated with laser currents in different ranges (38, 42,
  39, 41, and 41 subjects in each group). Dashed lines and bands of
  the corresponding color show means and standard deviations of the
  mean (see {\em Materials and methods}) of fits to these empirical
  data by the chosen model. Only the velocity in the range of time
  $[-1,2.25]$ s relative to the time of the onset of the laser
  stimulus was used for fitting. While most worms are still moving
  backward during this range of time, the inferred model {\em
    predicts}, without any additional free parameters, the time at
  which a worm's speed again becomes positive both as a function of
  (B) applied laser current and (C) peak observed worm reverse
  speed. These predictions (red curves) agree well with the binned
  averages (orange points with error bars representing the standard
  error of the mean if the bin has more than five worms) of individual
  worms' behavior (blue dots).\label{binnedOutput}%
}
\end{figure*}

\subsection*{Experimental model system} 

Nociception evokes a rapid escape behavior designed to protect the
animal from potential harm~\cite{eaton2013neural,Pirri:2012aa}. {\em
  C.\ elegans}, a small nematode with a simple nervous system, is a
classic model organism used in the study of nociception. A variety of
studies have used {\em C.\ elegans} to elucidate genes and neurons
mediating nociception to a variety of aversive stimuli including high
osmolarity and mechanical, chemical, and thermal
stimuli~\cite{Bargmann01011990,Hilliard63,Kaplan2227,Wittenburg1999}.
However, a complete dynamic understanding of the escape response at the
neuronal, let alone the molecular, level is not fully known. Recent
studies have quantified the behavioral escape response of the worm
when thermally stimulated with laser
heating~\cite{Mohammadi:2013ku,LeuMohRyu16}, and these data will be
the focus of our study. The response is dynamic: when the stimulus is
applied to the animal's head, it quickly withdraws, briefly
accelerating backwards, and eventually returns to forward motion,
usually in a different direction. Various features of this response
change with the level of laser heating, such as the length of time
moving in the reverse direction and the maximum speed attained.

\subsection*{Fits and Predictions} 

We use the worm center-of-mass speed, $v$, as the variable whose
dynamics needs to be explained in response to the laser heating pulse.
We define $v>0$ as the worm crawling forward and $v<0$ as the worm
retreating backwards. The input to the model is the underlying
temperature, $h(t)$, which can be approximated as $h(t)=Ih_0(t)$,
where $I$ is the experimentally controlled laser current, and $h_0$ is
the temperature template, described in {\em Materials and Methods}.
Based on trajectories of 201 worms in response to laser currents
ranging between 9.6 and 177.4 mA, we let {\em Sir
  Isaac} determine the most likely dynamical system explaining this
data within the sigmoidal networks model class \cite{DanNem15b} (see
{\em
  Materials and Methods} for a detailed description of the modeling
and inference). The inferred model has a latent (unobserved) dynamical
variable, hereafter referred to as $x_2$, in addition to the speed.
$v$ and $x_2$ are coupled by nonlinear interactions.
However, some of these nonlinear interactions may be insignificant,
and may be present simply because the nested hierarchy introduces them
before some other interaction terms that are necessary to explain the data. Thus we reduce the model by setting parameters that are small to zero one by one and in various combinations, refitting such reduced
models, and using Bayesian Model Selection to choose between the
reduced model and the original {\em Sir Isaac} inferred model. The
resulting model is
\begin{align}
\label{reduced1}
  \frac{dv}{dt} &= -\frac{v}{\tau_1}
                  + V_1 h(t) 
    + \frac{W_{11}}{1+e^{v+ \theta_1}}
    + \frac{W_{12}}{1+e^{x_2}}, \\ \label{reduced2}
\frac{d x_2}{dt} &=  -x_2
	 + V_2 \, h(t),\quad x_2(t=0)=0.
\end{align}
Here $V_1,V_2,W_{11},W_{12}$ are constants inferred from data, while
the model uses the default value of $1.0$ s for the characteristic
time scale of the dynamics of $x_2$ (see {\em Materials and Methods}
for the values of the parameters). Interestingly, the inferred model
reveals that the latent dynamical variable $x_2$ is a linear low-pass
filtered (integrated) version of the heat signal.

The fits produced by this model are compared to data in
Fig.~\ref{binnedOutput}(A), showing an excellent agreement (see {\em
  Materials and Methods} for quantification of the quality of fits). Surprisingly, the quality of the fit for this automatically
generated model is {\em better} than that of a manually curated model
\cite{LeuMohRyu16}: only about 10\% of explainable variance in the
data remains unexplained by the model for times between 100 ms and 2 s
after the stimulus, compared to about 20\% for the manual model,
cf.~Fig.~\ref{fig:modelquality}(A).

However, the quality of the fit is not surprising in itself since the
{\em Sir Isaac} model hierarchy can fit any dynamics using sufficient data. A utility of a mathematical model is in its ability
to make {\em predictions} about data that were not used in
fitting. Thus we use the inferred model to predict when the
worm will return to forward motion, which usually happens well
after the temporal range used for fitting.
Figure~\ref{binnedOutput}(B,C) compares these predictions with
experiments, showing very good predictions. Such ability
to extrapolate beyond the training range is usually an indication that
the model captures the underlying physics, and is not purely
statistical \cite{nelson}, giving us confidence in using the model for
inferences about the worm.


\begin{figure*}[t]
\centering
\includegraphics[width=6.5in]{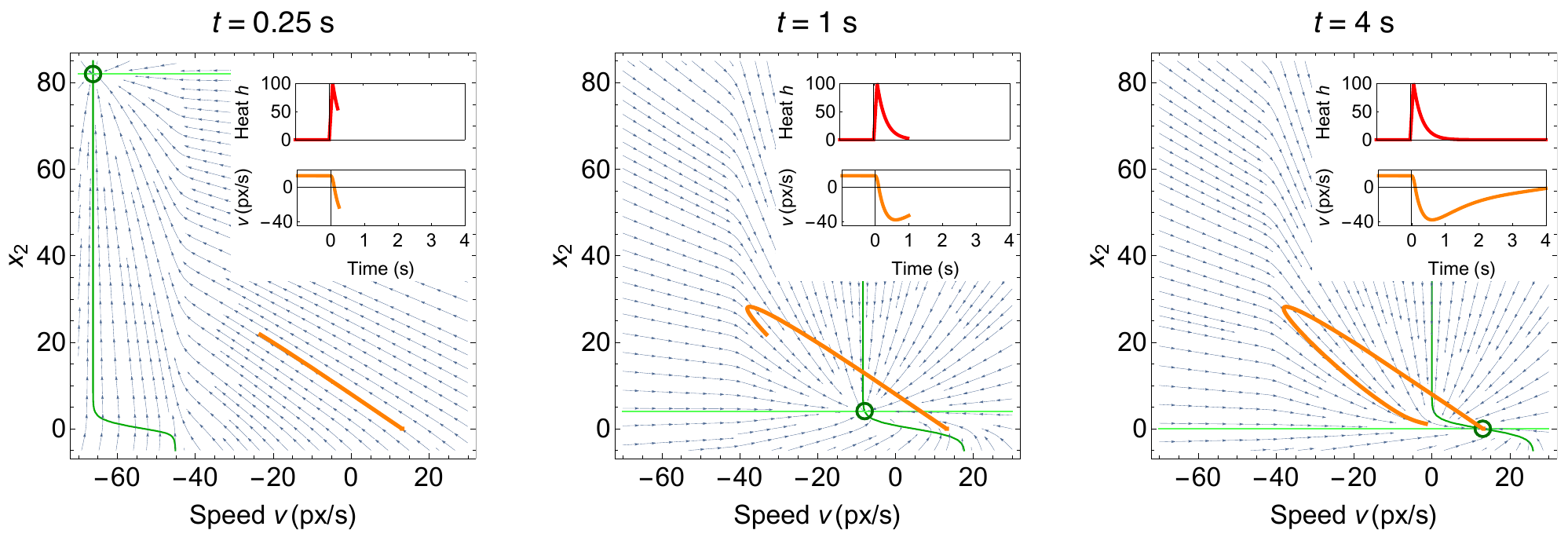}\textit{}
\caption{ \textbf{Phase space structure of inferred model.} With one
  hidden variable $x_2$, the model dynamics can be visualized in the
  two-dimensional phase space. As the instantaneous heat input $h$
  returns to zero after a brief pulse (red curve in insets), a single
  fixed point in the two-dimensional $(v,x_2)$ dynamics moves from
  negative speed (escape) to positive (forward motion). As the
  velocity of the worm trails the fixed point, this produces first a
  fast
  escape and then a slow return to forward motion (speed trajectory
  in orange in inset and in the phase portrait plots). Blue arrows
  indicate flow lines, circles indicate stable fixed points, and green
  lines indicate nullclines (dark green where $dv/dt = 0$ and light
  green where $dx_2/dt = 0$). \label{phaseSpaceAllTrials}%
} \end{figure*}

\subsection*{Model analysis} 
The algorithm has chosen to include a single latent dynamical
variable, which is a linear leaky integrator of the experienced
temperature. Having access to both the instantaneous stimulus and its
integral over the immediate past allows the worm to estimate the rate
of change in the stimulus. This agrees with the observation
\cite{Mohammadi:2013ku} that both the current temperature, as well as
the rate of its increase, are noxious for the worm. From this, one
could have guessed, perhaps, that {\em at least} one latent variable
(temperature derivative, or temperature at some previous time) is
required to properly model the escape response. However, the fact that
{\em Sir Isaac} inferred this from time series data alone and was able
to model the data with {\em exactly} one hidden variable is
surprising.

Figure~\ref{phaseSpaceAllTrials} shows the phase portraits of the
inferred dynamical model, Eqs.~(\ref{reduced1}, \ref{reduced2}), as well as the dynamics
of the speed and the heat stimulus $h$. Crucially, we see that there
is only one fixed point in the phase space at any instant of time, and
the position of this fixed point is affected by the current laser
stimulus value. This suggests that, at least at the level of the
population-averaged response, the behavior does not involve switching
among alternative behaviors defined dynamically as multiple fixed points
or limit cycles (e.~g., forward and backward motion) with the
switching probability influenced by the stimulus \cite{Stephens2008}, but rather the
stimulus controls the direction and the speed of the single dominant
crawling state.

\begin{figure}[t]
\centering
\includegraphics[width=3in]{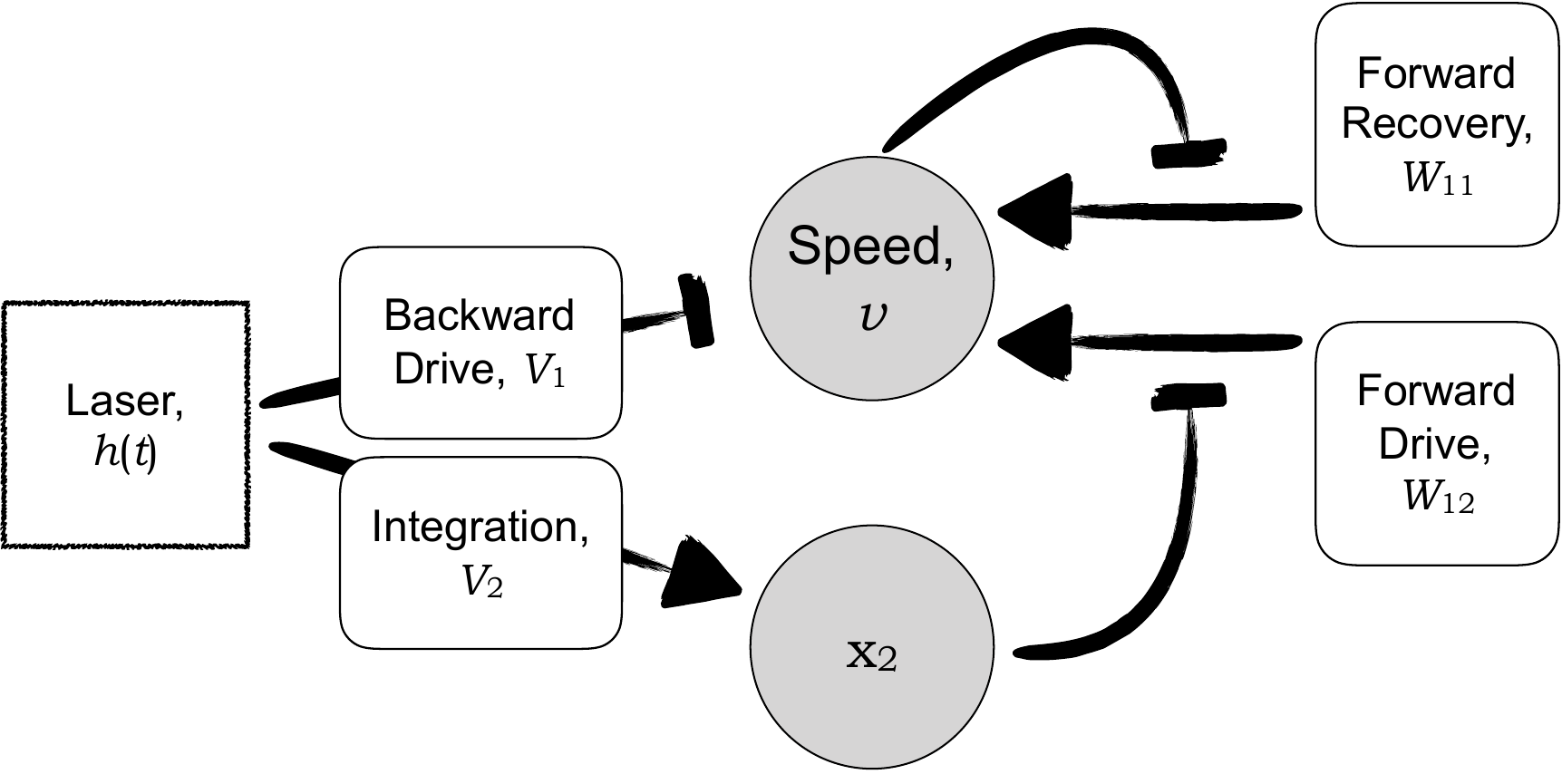}
\caption{{\bf Network diagram of the inferred model.} Variables,
  interactions, parameter values, and biological
  mechanisms are described in the main text.\label{fig:network}} 
\end{figure}

The network diagram of the model in Eqs.~(\ref{reduced1},
\ref{reduced2}) is shown in Fig.~\ref{fig:network}, where we omit the
linear degradation terms for $v$ and $x_2$. With the maximum
likelihood parameter values, Tbl.~\ref{fitParameterTableReduced}, the
model can be interpreted as follows. The hidden variable $x_2$ is a
linear leaky integrator of the heat signal, storing the average recent
value of the stimulus over about 1 s. While we do not know which exact
neuron can be identified with $x_2$, the thermosensory neurons AFD,
FLP located in head of the worm are strong candidates
\cite{Goodman2018,Liu2012}. The thermosensory neurons AFD respond to
changes in temperature and are the primary sensors responsible for
thermotaxis \cite{Goodman2018}. The sensory neuron FLP also is
thermosensitive and has a role in the thermal sensory escape response
\cite{Liu2012}.  When $x_2$ is near zero, the $W_{12}$ term is large, and,
together with the linear relaxation of the speed, $-v/\tau_1$, it
establishes a constant positive forward motion. We identify this term
with the forward drive command interneurons AVB, PVC
\cite{White1986}. After the temperature increases, $x_2$ grows.  It
rapidly increases the denominator in the $W_{12}$ term and hence shuts
down the forward drive. This is again consistent with the literature
indicating that the worms pause with even reasonably small temperature
perturbations \cite{Mohammadi:2013ku}.  An additional effect of the
stimulus is to directly inject a negative drive $-V_1h$ into the
dynamics of the velocity. When the stimulus is large, the $V_1$ term
is sufficiently negative to result in the velocity overshooting the
pause into the negative, escape range. We identify the $V_1$ term with
the reverse command interneurons AVA, AVD, and AVE, activated by the
thermosensory neurons AFD, FLP \cite{White1986}. When
$v>-\theta_1\approx -42$ px/s, the $W_{11}$ term is
suppressed. However, during fast escapes this suppression is lifted,
activating the positive drive $W_{11}$, which leads to faster recovery
of the forward velocity.  We identify the $W_{11}$ term with internal
recovery dynamics of the reverse command interneurons whose molecular
mechanisms of activity are only partially understood \cite{Gao2015}.
The velocity does not just relax to zero over some characteristic
time, but crosses back into the forward crawl once $x_2$ has decreased
sufficiently to reactivate the $W_{12}$ term.  Overall, the biological
interpretability of the model is striking. And where there is no
direct match between known worm biology and the model, the model
strongly suggests that we should be looking for specific predicted
features, such as the neural and molecular mechanisms for both
sensing the heat stimulus and its recent average.

Another notable feature of the network diagram is that it is similar
to other well-known sensory networks,  namely chemotaxis
(and the related thermotaxis) in {\em E.\
  coli} \cite{Sourjik2012,Paulick2017} and chemotaxis in {\em D.\
  discoideum} \cite{Jilkine:2011fy, Levchenko:2002cg}. In all these cases, the current value of the stimulus
is sensed in parallel to the stimulus integrated over the recent time.
They are later brought together in a negative feedback loop ({\em E.\
  coli}) or an incoherent feedforward loop ({\em D.\ discoideum}), 
  resulting in various adaptive behaviors. In contrast,
the worm's behavior is more complicated: the ambient temperature
participates in a negative feedback loop on the speed through
$W_{11}$, which results in adaptation. At the same time, the
integrated temperature (through $W_{12}$) and the current temperature
add {\em coherently} to cause the escape in response to both the
temperature and its rate of change. This illustrates the
difference between sensing, when an organism needs to respond to stimulus
changes only, and escape, where more complicated
dynamics is needed. 

\section*{Discussion}

In this article, we have used modern machine learning to
learn the dynamics underlying the temperature escape behavior in {\em
  C.~elegans}. The resulting automatically inferred model is more
accurate than the model curated by hand. It uses the dynamics within what is normally considered as discrete behavioral states to make extremely
precise verifiable (and verified) predictions about the behavior of
the worm beyond the
  range of time used for training. The model is fully interpretable,
with many of its features having direct biological, mechanistic
equivalents. Where such biological equivalents are unknown, the model
makes strong predictions of what they should be, and suggests what
future experiments need to search for.

One can question if describing the {\em C.~elegans} nociceptive
behavior, which typically is viewed as stochastic
\cite{Mohammadi:2013ku,LeuMohRyu16} and switching between discrete
states, with the deterministic dynamics approach of {\em Sir Isaac} is
appropriate. The quality of the fits and predictions is an indication
that it is. This is likely because (i) the escape is, to a large
extent, deterministic, becoming more and more so as the stimulus
intensity increases \cite{Mohammadi:2013ku}; (ii) on the scale of
individual worms, the discretized behavior states (forward, backward,
pause, \dots) have their own internal dynamics, with different
time-dependent velocities, which {\em Sir Isaac} models well, and the
boundaries between the states are not highly pronounced; and (iii) our
equations model a population of worms, and so even if individual
worms were dominated by stochasticity, {\em Sir Isaac} would do just
fine modeling the dynamics of the mean  behavior.

Crucially, the model discovers that the behavior, at least of an
average worm,  is not a simple
one-to-one mapping of the input signal: the instantaneous stimulus and
its temporally integrated history (one latent variable) are both
important for driving the behavior. The behavior is driven by one fixed point in the velocity-memory
phase space, and the worm changes its speed while chasing this fixed
point, which in turn changes in response to the stimulus. This is in
contrast to other possibilities, such as the worm being able to exhibit
both the forward and the backward motion at any stimulus value, and the 
stimulus and its history merely affect the probability of engaging in 
either of these two behaviors.

This is the first successful application of automatic phenomenological inference to modeling dynamics of complex biological systems, without using the helping constraints imposed by (partial) knowledge of the underlying biology. The emphasis on interpretable, physical models allows extrapolation well beyond the data used for training, which is hard for purely statistical methods. The automation allows for a comprehensive search through the model space, so that the automatically inferred model is better than the human-assembled one, especially when faced with only partial knowledge of the complex system. Even for the best studied biological systems, we do not have the necessary set of measurements to model them from the ground up. Our work illustrates the power of phenomenological modeling approach, which allows for top down modeling, adding interpretable constraints to our understanding of the system.

\section*{Materials and Methods}

\subsection*{Data collection}

A detailed description of the experimental methods has been previously
published \cite{LeuMohRyu16}. In summary, we raised wild-type, N2 {\em
  C.~elegans} using standard methods, incubated at 20$^\circ$C with
food. Individual worms were washed to remove traces of food and placed
on the surface of an agar plate for 30 minutes at 20$^\circ$C to
acclimatize. Worms were then transferred to an agar assay plate seeded
with bacteria (food) and left to acclimatize for 30 more minutes. The
worms were then stimulated with an infra-red laser focused to a
diffraction limited spot directed at the ``nose'' of the worm. The
intensity of the stimulus was randomized by selecting a laser current
between 0 and 200 mA, with a duration of 0.1s. The temperature
increase caused by laser heating was nearly instantaneous and reached
up to a maximum of 2$^\circ$C for our current range.  Each worm was
stimulated only once and then discarded. Video of each worm's escape
response was recorded at 60 Hz and processed offline using custom
programs in LabVIEW and MATLAB.

\subsection*{Input data}

Data used for dynamical inference are as described in
Ref.~\cite{LeuMohRyu16}. Speed data for 201 worms were extracted from
video frames at 60 Hz and smoothed using a Gaussian kernel of width
500 ms. We use data between 0.5~s before and 2.25~s after the start of
the laser stimulus. Aligning the data by laser start time, the
stimulus happens at the same time in each trial. Naively used, this
can produce models that simply encode a short delay followed by an
escape, without requiring the stimulus input. To ensure that instead
the stimulus causes the response in the model, for each trial we add a
random delay between 0 and 1 s to the time data.

Additionally, we are not interested in capturing any dynamics in the
pre-laser free crawling state. If we only use the small amount of 
pre-laser data measured in this experiment, the inference procedure 
is free to include models with
complicated transient behavior before the stimulus. For this reason,
we include a copy of pre-laser data at a fictitious ``equilibration time'' long before
the stimulus time (10 s), artificially forcing the model to develop a
pre-stimulus steady state of forward motion. Finally, we weight the
pre-laser data such that it appears with equal frequency as post-laser
data in fitting, in order that the inference algorithm is not biased toward
capturing post-laser behavior more accurately than the pre-laser one.

\subsection*{Estimating explainable variance}

The observed variance in worm speed can be partitioned into that
caused by the input (changes in laser current) and that caused by
other factors (individual variability, experimental noise, etc.). As
in Ref.~\cite{LeuMohRyu16}, we focus on our model's ability to capture
the former, ``explainable'' variance. We treat the latter variance as
``unexplainable'' by our model, and it is this variance that we use to
define uncertainties on datapoints for use in the inference procedure.
We estimate these variances by splitting the data into trials with
similar laser current $I_\mu$ (5 bins), producing variances $\sigma_{\mu}^2(t)$
that depend on the laser current bin and on the time relative to
laser start. For simplicity, we use a constant uncertainty for all
datapoints that is an average over laser current bins and
times relative to the stimulus onset, which is equal to $\sigma = 14.2$ px/s. We use this uncertainty when
calculating the $\chi^2$ that defines the model's goodness-of-fit:
\begin{equation} \label{chiSq} \chi^2(\Theta) = \sum_{i=1}^{N_D}
  \left(\frac{x_i^\mathrm{model}(\Theta) - x_i^\mathrm{data}}{\sigma}
  \right)^2, \end{equation}
where the index $i$ enumerates data points used for the evaluation, $N_D$ is the number of points, and $\Theta$ is the vector of all parameters.

\begin{figure}[t]
\centering
\includegraphics[width=2.5in]{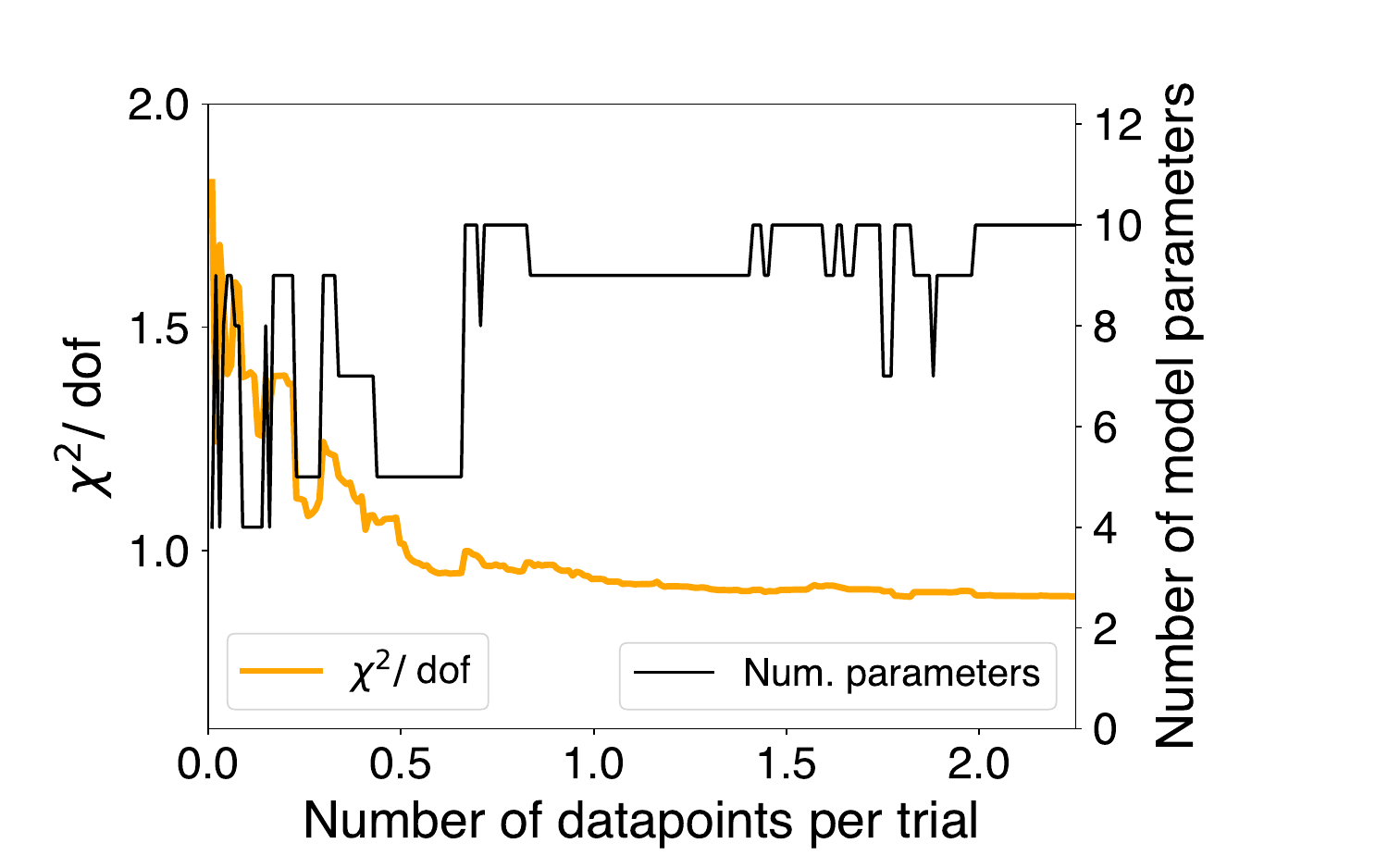}
\caption{ 
  \textbf{Monitoring goodness of fit in the process of model
    inference.} {\em Sir Isaac}  adds data gradually to aid in
  parameter fitting. As data is added, the selected model includes
  more detail (number of model parameters in black) until it saturates
  to the 10 parameter model we use. The goodness of fit ($\chi^2$ per
  degree of freedom, in yellow) is measured using all data, including
  time points not used in model fitting. \label{goodnessOfFit}%
} \end{figure}

\subsection*{{\em Sir Isaac} inference algorithm}

We use the {\em Sir Isaac} dynamical inference algorithm \cite{DanNem15b} to
find a set of ODEs that best describes the data without overfitting.
Based on previous studies of simulated biological systems, 
we use the continuous-time 
sigmoidal network model class,
which produces a set of $J$
ODEs of the form
\begin{equation}
\frac{d x_i}{dt} = -x_i/\tau_i
	+ V_{i} \, h(t) + \sum_{j=1}^{J} W_{ij} \, \xi(x_j + \theta_j),
\end{equation}
where $\xi(y) = 1/(1+\exp (y))$ and $h(t)$ is the sensory input defined
in \eqref{sensoryInput}. The algorithm infers both the number of
parameters (controlled by the total number of dynamical variables $J$)
and the parameter values themselves (the timescales $\tau_i$,
interaction strengths $V_i$ and $W_{ij}$, and biases $\theta_j$). The
first dynamical variable $x_1$ is taken to be the signed speed of the
worm's center of mass, in pixels per second, with negative
values corresponding to backward motion. Further dynamical variables
$x_i$ with $i>1$ correspond to latent (unmeasured) dynamical
variables.

The fitting procedure starts by using one datapoint (one random time)
from each of a few trials, then gradually adds trials and eventually
multiple datapoints per trial, refitting model parameters at each step.
The resulting model fits are scored based on their performance in
predicting the entire time series (see \figref{goodnessOfFit} for the fit quality). When the performance and model
complexity of the winning model saturate, we use the resulting model
as our description of the system. In this way, parameters are fit
using only a small subset of the available data---we find that using
$\sim2$ randomly chosen timepoints per trial (out of the total 165) is
sufficient, cf.~\figref{goodnessOfFit}.  This approach significantly
reduces computational effort (which scales linearly in the number of
datapoints used) and minimizes the effects of correlations between
datapoints that are close to one another in time.  Finally, it
prevents the optimization from getting stuck at local minima and
saddles, which change as new data points are added, or data is
randomized.  These reasons are similar to the reasons behind
stochastic gradient descent approaches
\cite{Mehta:2018tt}. The developed software is available from \url{https://github.com/EmoryUniversityTheoreticalBiophysics/SirIsaac}.

\begin{table}[t]
\caption{\label{paramsTable} Parameters used for input data, dynamical model inference, 
  and the dynamics of the sensory input. The inference parameters quoted are used in the {\em Sir Isaac} 	      	algorithm: ``complexity stepsize'' sets which models in the model hierarchy are
  tested, with 1 indicating that no models are skipped as parameters
  are added to make models more complex; here ``ensemble'' refers to
  the parameter ensemble used to avoid local minima during parameter
  fitting; ``avegtol'' and ``maxiter'' are parameters controlling the
  local minimization phase of parameter fitting; ``stopfittingN'' sets
  the number of higher-complexity models needed to be shown to have
  poorer performance before selecting a given model and finishing the optimization; ``connection
  order'' and ``type order'' set the order of adding parameters within
  the model hierarchy \cite{DanNem15}; and ``prior $\sigma$'' is the
  standard deviation of the Gaussian priors on all parameters.}%
\begin{tabular}{llll}
     \hline
                                 \multicolumn{2}{l}{{\bf Input data parameters}} \\
                                 equilibration time					& 10 s		\\
                                 \hline
                                  \multicolumn{2}{l}{{\bf Inference parameters}}  \\
                                 complexity stepsize			& 1			&	
                                 tested ensemble members & 10   \\ 
                                 maxiter   & 100 &
                                 ensemble temperature 		& 100 				\\
                                 avegtol 	   				& $10^{-2}$		&	
                                 ensemble generation steps   & 1000		\\
                                 stopfittingN				& 5 				&
                                 connection order			& `node' 			\\
                                 type order					& `last' 		&	
                                 prior $\sigma$				& 100 				\\
                                 \hline
                                  \multicolumn{2}{l}{{\bf Sensory input parameters}}  \\
                                 $\tau_{\mathrm{on}}$		& 0.1 s 			&
                                 $\tau_{\mathrm{decay}}$
                                                                                                & 0.25 s			\\
                                \hline
                               \end{tabular}
\end{table}

\subsection*{Inferred models}

The inference procedure produces the following
differential equations:
\begin{align}
\label{inferredModelEquations1}
\frac{d v}{dt} &= -\frac{v}{\tau_1}
	 + V_1 \, h(t) 
    + \frac{W_{11}}{1+\exp(\mathrm{speed} + \theta_1)} 
                 + \frac{W_{12}}{1+\exp(x_2)}, \\ \label{inferredModelEquations2}
  \frac{d x_2}{dt} &= -x_2
                     + V_2 \, h(t) 
    + \frac{W_{21}}{1+\exp(\mathrm{speed} + \theta_1)}
    + \frac{W_{22}}{1+\exp(x_2)}.
\end{align}
The model includes one latent dynamical variable $x_2$. The maximum
likelihood fit parameters are shown in Table~\ref{fitParameterTable}.
(Note that the selected model did not include variables $\tau_2$ and
$\theta_2$, so they are set to their default values: $\tau_2=1$ and
$\theta_2=0$.)

We notice that some of the inferred parameters are close to zero, and so we check whether the model can be simplified by setting each of them to zero one at a time and in combinations and then measuring the approximate Bayesian model selection posterior likelihood
score \cite{DanNem15b} for the original {\em Sir Isaac} inferred model and for each of the reduced models. The original model has the log-likelihood of $-197.2$, and the best model, Eqs.~(\ref{reduced1}, \ref{reduced2}), has the highest Bayesian likelihood of $-192.9$.  This becomes our chosen model, maximum likelihood parameters for which
can be found in Table~\ref{fitParameterTableReduced}.

\begin{table}[t] 
\caption{Maximum-likelihood parameters for the model inferred by {\em
    Sir Isaac}. We do not calculate parameter co-variances 
  since posteriors are sloppy \cite{Transtrum:2015hm} and
  non-Gaussian around the maximum likelihood. Instead we
  estimate the standard
  deviation of the fits themselves,  cf.~Fig.~\ref{binnedOutput}(A). Parameter values are reported
  to an accuracy of about one digit beyond the least statistically
  significant one. \label{fitParameterTable} }%
\begin{tabular}{ c  l c l } 
\hline 
$\tau_1$ & $1.33$ s & 
$\theta_1$ & $42.1$ px/s  \\ 
$V_1$ & $-2.30$ px/(s$^2$ mA) & 
$V_2$ & $1.7$ 1/(s mA) \\ 
$W_{11}$ & $81.0$ px/s$^2$ & 
$W_{12}$ & $44.1$ px/s$^2$ \\ 
$W_{21}$ & $-0.1$ s$^{-1}$ & 
$W_{22}$ & $5.5$ s$^{-1}$ \\ 
$v(t=0)$ & $12.6$ px/s  & 
$x_2(t=0)$ & $1.6$ \\ 
\hline 
\end{tabular} 
\end{table}

\begin{table}[t] 
\caption{Maximum-likelihood parameters for the reduced model. \label{fitParameterTableReduced}}%
\begin{tabular}{ c  l  c l } 
\hline 
$\tau_1$ & $1.33$ s & 
$\theta_1$ & $42.1$ px/s  \\ 
$V_1$ & $-2.30$ px/(s$^2$ mA) & 
$V_2$ & $1.5$ 1/(s mA) \\ 
$W_{11}$ & $76.$ px/s$^2$ & 
$W_{12}$ & $19.7$ px/s$^2$ \\ 
$v(t=0)$ & $12.1$ px/s  & 
\\ 
\hline 
\end{tabular} 
\end{table}

\begin{figure*}[t] \noindent\centering
    \includegraphics[height=1.7in]{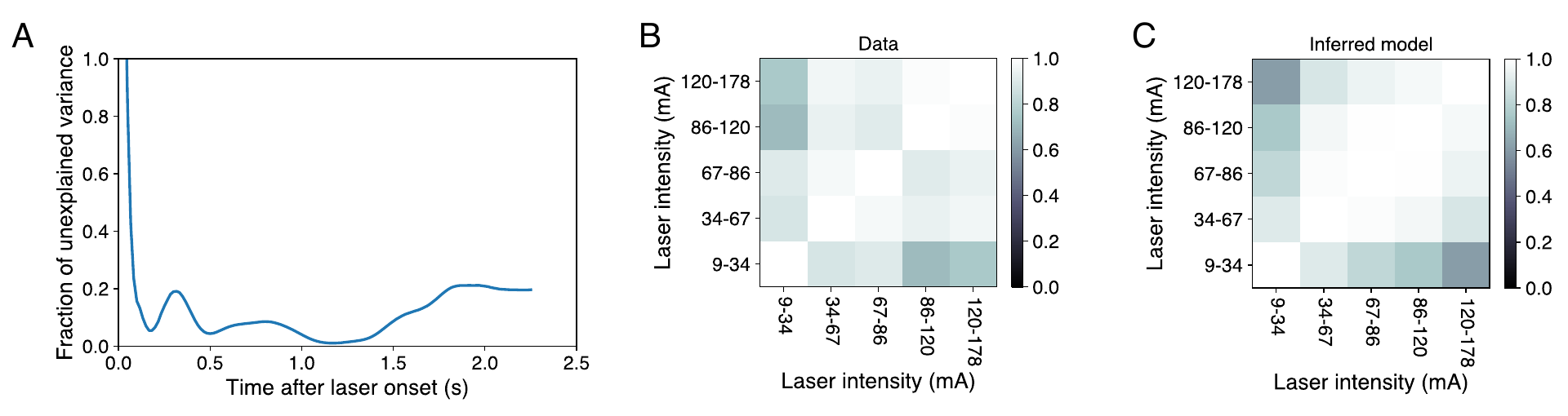}
  \caption{\label{fig:modelquality}{\bf Comparison of the automated
      model to the manually curated model of Ref.~\cite{LeuMohRyu16}.} (A)
    The fraction of the overall explainable velocity variance not
    explained by our model; the unexplained variance is roughly half
    of that of Ref.~\cite{LeuMohRyu16}. (B, C) The escape response has been
    modeled previously as stereotypical, with an overall scaling that
    is laser current dependent (for laser currents $>25$ mA). Perfect
    stimulus-dependent rescaling of the escape would correspond to a
    correlation of 1 between velocity traces for different stimuli. (B)
    shows these correlations, over time, between mean response speeds
    in different laser current bins. (C) shows the same correlations
    in the inferred model. To the extent that correlations in both
    panels are nearly the same, our model
    recovers the approximate stereotypical nature of the response ~\cite{LeuMohRyu16}.} \end{figure*}
\subsection*{Bayesian model of uncertainty}

To quantify uncertainty in model structure and parameters, we take a
Bayesian approach and sample from the posterior distribution over
parameters. Assuming independent Gaussian fluctuations in the data
used in fits, the posterior is simply $\exp{(-\chi^2(\Theta))}$, with
$\chi^2(\Theta)$ from \eqref{chiSq}. We use a standard Metropolis
Monte Carlo sampler, as implemented in SloppyCell \cite{sloppy}, to
take $100$ samples (each separated by $1000$ Monte Carlo steps) used
to quantify uncertainty in the fit, cf.~Fig~\ref{binnedOutput}(A).

\subsection*{Model of sensory input}

Each experimental trial begins with a forward moving worm, which is
stimulated with a laser pulse of duration $\tau_{on} = 0.1$ s starting
at time $t=0$. The worm's nose experiences a quick local temperature increase 
$h(t)$, which we model as a linear increase during the  pulse,
with slope proportional to the laser current $I$. The time scale of the temperature decay of the heated area to the ambient temperature due to heat diffusion is $0.15$ s  \cite{Mohammadi2013}. However, the stimulation area is broad ($220 \mu$m, FWHM), and as the worm retreats, its head with the sensory neurons first moves deeper into the heated area, before the temperature eventually decreases. Thus the dynamics of the sensory input is complex and multiscale. However, since each individual behaves differently, and we do do not measure individual head temperatures, we model the {\em average} sensory stimulus past the heating period as a single exponential decay with the longer time scale of $\tau_{\rm decay}=0.25$ s:
\begin{equation} \label{sensoryInput}
  h(t)/\alpha = \begin{cases}
      0 & t \leq 0, \\
      I t / \tau_{\mathrm{on}} & 0 \leq t \leq \tau_{\mathrm{on}}, \\
      I \exp{[-(t-\tau_{\mathrm{on}})/\tau_{\mathrm{decay}}]} & t \geq \tau_{\mathrm{on}}.
   \end{cases}
\end{equation}
Here $\alpha$ has units
temperature per unit of laser current. For convenience and without loss of
generality, we set $\alpha = 1$ and absorb its definition into the
$V_i$ parameters that multiply $h$ in  Eqs.~(\ref{inferredModelEquations1}, \ref{inferredModelEquations2})
giving $V_i$ units of the time derivative of $x_i$ per unit current.

\subsection*{Comparison to the model of Ref.~\cite{LeuMohRyu16}}

A quantitative model of {\em C.~elegans} nociceptive
response was constructed in Ref.~\cite{LeuMohRyu16}. It involved
partitioning worms into actively escaping and (nearly) pausing after a
laser stimulation, estimating the probability of pausing as a function
of the applied laser current, and finally noticing that the mean
response of the active worms is nearly stereotypical, with the
response amplitude depending nonlinearly on the stimulus
intensity. The same stimulus  causes varied response
trajectories in individual worms, and this variability is unexplainable
in any model that only relates the stimulus to the population averaged
response velocity. Of the variability in the response that is explainable
by the stimulus, the model of Ref.~\cite{LeuMohRyu16} could not account for $\sim20\%$ in
the range from a few hundred ms to about 2 s post-stimulation. Figure
\ref{fig:modelquality}(A) shows that {\em Sir Isaac} our selected 
model captures {\em more} of the data variance, leaving only 
$~10\%$ of the explainable variance unexplained over much of the same
time range. Figure \ref{fig:modelquality}(B,C) illustrates that {\em Sir
Isaac} has recovered the approximate stereotypy in the response to the
same accuracy as it is present in the data.


\section*{Acknowledgments}
This work was partially supported by NIH Grants EB022872 and NS084844, James S.\ McDonnell Foundation Grant
  220020321, NSERC Discovery Grant, and NSF Grant IOS-1456912. We are
  grateful to the NVIDIA corporation for 
  donated Tesla K40 GPUs. We are grateful to the Aspen Center for
  Physics, supported by NSF grant PHY-1607611, for supporting this research within ``Physics of Behavior'' programs.

\bibliography{worm}

\end{document}